\documentstyle[floats,twocolumn,prl,aps,psfig]{revtex}
\newcommand{\be}{\begin{equation}}
\newcommand{\ee}{\end{equation}}
\newcommand{\bea}{\begin{eqnarray}}
\newcommand{\eea}{\end{eqnarray}}
\newcommand{\les}{\ell_{\hbox{\tiny ES}}}
\newcommand{\oni}{\omega_{\hbox{\tiny NI}}}
\newcommand{\omf}{\omega_{\hbox{\tiny MF}}}
\begin{document}
\draft

\title{Spatio-temporal distribution of nucleation events
during crystal growth}

\author{Claudio Castellano$^{1,*}$ and  Paolo Politi$^{2,\dag}$}
\address{$^1$ Istituto Nazionale per la Fisica della Materia, 
Unit\`a di Roma ``La Sapienza", P.le Aldo Moro 2, 00185 Roma, Italy}
\address{$^2$ Istituto Nazionale per la Fisica della Materia, Unit\`a di
Firenze, L.go E. Fermi 2, 50125 Firenze, Italy}

\author{(\today)}
\author{\parbox{397pt}{\vglue 0.3cm \small
We consider irreversible second--layer nucleation that
occurs when two adatoms on a terrace meet. We solve the problem
analytically in one dimension for zero and infinite step--edge
barriers, and numerically for any value of the barriers
in one and two dimensions.
For large barriers, the spatial distribution of nucleation events 
strongly differs from $\rho^2$, where $\rho$ is the stationary adatom density
in the presence of a constant flux.
The probability $Q(t)$ that nucleation occurs at time $t$ after
the deposition of the second adatom,
decays for short time as a power law [$Q(t)\sim t^{-1/2}$] in $d=1$ and 
logarithmically [$Q(t)\sim 1/\ln(t/t_0)$] in $d=2$; 
for long time it decays exponentially.
Theories of the nucleation rate $\omega$ based on the assumption that
it is proportional to $\rho^2$ are shown to overestimate $\omega$ 
by a factor proportional to the number of times an adatom diffusing on
the terrace visits an already visited lattice site.
}}
\maketitle

Molecular Beam Epitaxy (MBE) is one of the most common techniques for 
growing nanoscale materials~\cite{libroMBE}: 
a controlled flux of particles
arrives ballistically on a substrate whose temperature is generally
high enough to activate surface diffusion of newly deposited adatoms. 
In the absence of preexisting steps, i.e. for high-symmetry surfaces,
growth proceeds through the formation of stable dimers (or largest
nuclei, depending upon the temperature, the flux 
and surface symmetry~\cite{libroJV})
and the subsequent aggregation of diffusing particles.

Nucleation processes take place on a flat surface during the submonolayer
regime~\cite{Brune} and afterwards on islands. 
In this Letter we are interested in the latter case, where the process
of nucleation occurs as follows: 
adatoms are deposited at a rate $F$ per unit time and lattice
site and diffuse at a rate $D$. The typical time between two deposition 
events on a terrace of size $L$ is $\tau_{dep} = (FL^d)^{-1}$ and
the typical residence time of an adatom on the terrace  is~\cite{KPM}
$\tau_{res}\sim L(L+\alpha_d \les)/D$, where $\les$ is the
Ehrlich-Schwoebel (ES) length~\cite{EV} that measures the strength
of step-edge barriers~\cite{KE} hindering the descent of adatoms;
$\alpha_d$ is a numerical factor depending on dimension only.
A nucleation event takes place when a newly deposited adatom
finds the previous one still on the terrace {\it and} they meet before
descending, forming a stable dimer.
Once two adatoms are on the terrace at the same time a nucleation
event takes place, if it does, in a time of order $\tau_{tr}\sim L^2/D$,
the typical time required for a traversal of the terrace~\cite{KPM}.
Therefore the probability that a third atom influences the nucleation
process is of order $p_3 \sim \tau_{tr}/\tau_{dep}$, which is very small
in all physically relevant cases. More precisely, in $d=2$ 
$p_3 \sim L^4/\ell_{\hbox{\tiny D}}^6$, where
$\ell_{\hbox{\tiny D}} \sim(D/F)^{1/6}$ -- the so called
diffusion length~\cite{libroJV} -- is the typical size of a terrace 
when nucleation occurs.
Since $L \leq \ell_{\hbox{\tiny D}}$ and
$\ell_{\hbox{\tiny D}} \gg 1$ 
we conclude that $p_3 \ll 1$ and for irreversible nucleation only
processes involving two adatoms must be take into account.

The simplest theoretical treatment of nucleation phenomena in crystal growth
is based on the assumption~\cite{Venables} that the nucleation rate $\omega$
is proportional to the square of the stationary
adatom density $\rho$ in the presence of a constant flux $F$.
Such a mean-field treatment has been used to estimate physical parameters
of materials~\cite{TDT,altro} and as an ingredient in mesoscopic models of
epitaxial growth~\cite{Ratsch00}.
However, recent papers~\cite{KPM,maass} have shown (via Monte Carlo
simulations and scaling arguments~\cite{maass} or theoretically~\cite{KPM}) 
that the mean field prediction $\omega\sim\rho^2$ for the nucleation
rate is not correct in $d=2$ for large ES barriers.
This prompted us to reconsider the general validity of
mean-field theory. It is a priori not clear to what extent
two-particle properties as $\omega$ can be simply obtained from $\rho$,
a quantity describing a single particle on a terrace.
In this Letter we solve exactly the problem of irreversible nucleation
on a terrace in 1 and 2 dimensions for all values of the ES barrier.
We compute (analytically or numerically) the spatial and temporal
distributions of nucleation events and the total nucleation rate.
In this way we are able to assess when and why the mean-field 
approximation provides
sufficiently accurate results and by how much it fails.

Let us start with the case $\les=0$.
The discrete evolution equation for the probability $p_n(t)$ of finding a
single adatom in site $n$ at time $t$ is
\be
p_n(t+1) = {1\over 2} [p_{n+1}(t) + p_{n-1}(t)].
\label{evolution_1d}
\ee
The boundary conditions are $p_0(t)=p_{L+1}(t)=0$.
 
By looking for solutions of Eq.~(\ref{evolution_1d}) of the form
$p_n(t)=N_n T(t)$ one finds the general solution
\be
p_n(t)=\sum_{k=1}^L A_k \cos^t \left({k\Pi}\right)
\sin\left({n k\Pi}\right),
\label{sol1d}
\ee
with coefficients $A_k = {2\over L+1} \sum_{n=1}^L p_n(0) \sin(n k\Pi)$,
where $\Pi=\pi/(L+1)$.

Nucleation occurs when two adatoms wandering on the same terrace meet.
The one-dimensional diffusion of two particles can be reformulated by
taking the coordinates $m,n$ of the two walkers as the coordinates
of {\it one} walker in $d'=2$ moving on a square lattice $(L\times L)$.
The generalization of (\ref{evolution_1d}) to two dimensions is 
$p_{m,n}(t+1)= {1\over 4}[p_{m+1,n}(t)+p_{m-1,n}(t)+p_{m,n+1}(t)
+p_{m,n-1}(t)]$
with boundary conditions $p_{0,n}=p_{L+1,n}=p_{m,0}=p_{m,L+1}=0$.
Again, separating space and time variables one obtains the
most general solution
\bea
p_{m,n}(t) = \sum_{k,j=1}^L {B_{kj}\over 2^t}\left[
\cos(k\Pi) + \cos(j\Pi)\right]^t
&&\nonumber\\
\times\sin(m k\Pi) \sin(n j\Pi),&&
\label{general}
\eea
where the coefficients $B_{kj}$ are
\be
B_{kj} = \left({2\over L+1}\right)^2 \sum_{m,n=1}^L p_{m,n}(0)
\sin(m k\Pi) \sin(n j\Pi)~.
\label{Bkj}
\ee 

The initial condition $p_{m,n}(0) = {\hat p}_m {\hat p}_n$
encodes the information about the way particles are deposited.
Notice that $t=0$ indicates in Eqs.~(\ref{general},\ref{Bkj}) the time when
the second adatom arrives.
We consider adatoms landing on the terrace with spatially uniform
probability $p_n^U = 1/L$.
However, since arrivals on the terrace are not simultaneous,
the distribution of the first adatom has changed to $p_m(t')$
(Eq.~\ref{sol1d}) when the second one lands on the terrace. 
The nucleation process depends therefore on the precise interarrival time
$t'$, which is a poissonian random variable
$P_{\hbox{\tiny{ARR}}}(t') = (\tau_{dep})^{-1}\exp(-t'/\tau_{dep})$.

One must compute physical quantities $O$ by averaging over $t'$:
$O = \sum_{t'=0}^{\infty} P_{\hbox{\tiny{ARR}}}(t') O(t')$,
where $O(t')$ is computed using as initial distributions $p_{m}(t')$
for the first particle and $p_n^U = 1/L$ for the second.
If $O(t')$ is linear with respect to the initial distributions (as are all the
quantities considered below), the sum can be replaced
by a single computation of $O$ with the initial distribution of the
first adatom given by
${\hat p}_m = \sum_{t'=0}^{\infty} P_{\hbox{\tiny{ARR}}}(t') p_m(t')$.

The general solution can be found~\cite{lungo}, but in the limit 
$\tau_{res} \ll \tau_{dep}$, which is realistic for MBE,
one can consider $P_{\hbox{\tiny{ARR}}}(t')$ 
as a constant on the scale $\tau_{res}$ and
then ${\hat p}_m = \sum_{t'=0}^{\infty} p_m(t')$.
Therefore ${\hat p}_m$ is proportional to the normalized stationary solution
$p_m^S$ of (\ref{evolution_1d}) in the presence of a constant 
flux~\cite{nota1}: $p_m^S=6/[L(L+1)(L+2)] m(L+1-m)$; $p_m^S$ will be taken
as the distribution for the first particle for computing $P(n)$ and $Q(t)$.

So far we have not considered the interaction between adatoms, i.e. the
fact that when the two particles meet they stop diffusing.
An irreversible nucleation event generally occurs when two adatoms 
are on nearest neighbor sites. To allow an analytic treatment 
we consider here a nucleation event to occur when the adatoms are on
the {\it same} site.
When this occurs the dynamics stops, implying $p_{m,m}(t)=0$ for all $t>0$.
This boundary condition can be taken into account by the classical image
method~\cite{Fisher84,Redner99}: The initial condition is chosen to
be antisymmetric with respect to particles exchange $p_{m,n}(0)=-p_{n,m}(0)$.
This implies $B_{kj}=-B_{jk}$ and therefore $p_{m,n}(t)=-p_{n,m}(t)$. 
In this way the boundary condition $p_{m,m}(t)=0$ is obeyed for all $t$
because the two triangles $(m>n)$ and $(m<n)$ are dynamically disconnected.

When the antisymmetric initial condition is not imposed adatoms diffuse
without interacting: they do not feel each other even if they are on the
same site and they wander until they get off the terrace.
In the following we will speak of nucleations for these noninteracting
adatoms, intending that they ``nucleate'' when they are on the same site.
Clearly two noninteracting adatoms may give rise to several
``nucleation'' events before leaving the terrace.

\begin{figure}
\centerline{\psfig{figure=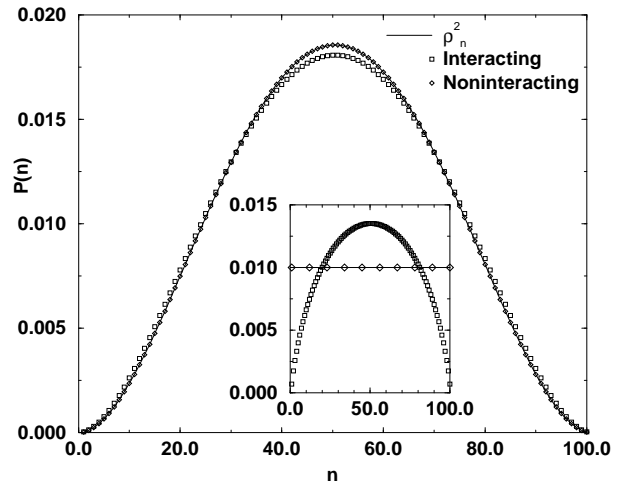,width=8cm,angle=-90}}
\caption{Main: normalized distribution of the nucleation sites $P(n)$ in $d=1$
for $\les=0$ and $L=100$.
$\rho_n \sim p_n^S$ is the (normalized) stationary solution of
Eq.~(\ref{evolution_1d}) in the presence of a constant flux. Inset:
The same for $\les=\infty$.}
\label{fig1}
\end{figure}

For noninteracting adatoms the coefficients $B_{kj}$ are simply
the product of the single adatom coefficients $A_k A_j$.
In the case of interacting adatoms we obtain~\cite{lungo}
$B_{kj} = \left[2/(L+1)\right]^2 [ B_{kj}^< - B_{jk}^< ],$
where
$ B_{kj}^< = \sum_{m<n}p_{m,n}(0) \sin (m k\Pi) \sin (n j\Pi)$.

For interacting adatoms the probability of a nucleation event on site $n$
at time $t+1$ is given by $(1/2)[p_{n,n+1}(t)+p_{n-1,n}(t)]$, while for
$t=0$ it is $p_{n,n}(0)$.
For noninteracting adatoms it is simply $p_{n,n}(t)$ at any time.
By summing over $t$ one obtains the spatial distribution $P(n)$ of nucleation
sites, which is reported in Fig.~\ref{fig1}.
It turns out immediately that the mean-field prediction $\rho_n^2$
is exact only if adatoms do not interact, i.e. nucleation events
following the first one are taken into account.
In the interacting case the distribution of nucleation sites 
differs from $\rho_n^2$, but the discrepancy is rather small.

We now discuss the distribution of nucleation times, i.e. the
probability $Q(t)$ that adatoms meet at time $t$ after the deposition
of the second adatom.

For non-interacting adatoms,
\be
Q(t)  =  \sum_{n=1}^L p_{n,n}(t)
= {L+1 \over 2} \sum_{k=1}^L B_{kk}
\cos^t(k \Pi)~.
\ee

We can rewrite $\cos^t(\phi)=\exp[t\ln\cos(\phi)]$ and since
the coefficients $B_{kk}$ diverge for small $k$ as $k^{-4}$, we expand
the cosine for small $\phi$ and finally extract the dominant contribution
coming from the mode $k=1$:
$Q(t) \sim \exp \left[-\Pi^2 t/2 \right]$.
\begin{figure}
\centerline{\psfig{figure=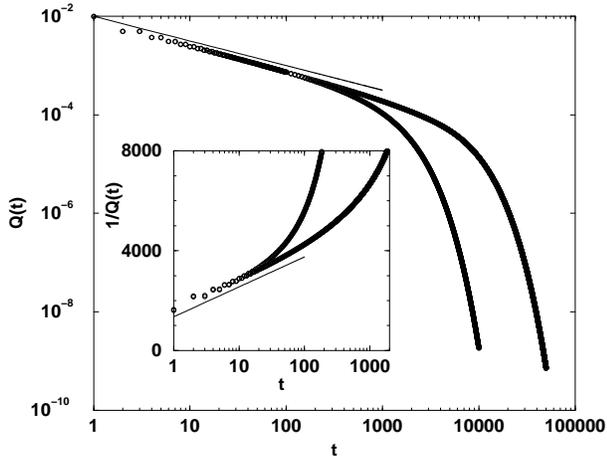,width=8cm,angle=-90}}
\caption{Main: $Q(t)$ on a log-log scale for $L=100$ in $d=1$
for $\les=0$ (bottom) and $\les=\infty$ (top). 
The straight line shows a decay of $Q(t)$ as $t^{-1/2}$. 
Inset: $1/Q(t)$ on a log-lin scale for $L=40$ in $d=2$
for $\les=0$ (top) and $\les=\infty$ (bottom).
The straight line shows a decay of $Q(t)$ as $1/\ln(t/t_0)$.
}
\label{fig2}
\end{figure}
In the case of interacting adatoms $Q(t+1) =
\sum_{n=1}^L {1 \over 2}[p_{n,n+1}(t) + p_{n-1,n}(t)]$.
By evaluating the summation, we obtain
$Q(t+1) = \sum_{k,j=1}^L B_{kj} C_{kj} 2^{-t}\left[
\cos(k \Pi) + \cos(j \Pi)\right]^t$,
where the \hbox{$C_{kj} = \sum_{n=1}^L
\sin[k \Pi (n-1)] \sin(j \Pi n)$} can be evaluated explicitly.
Approximations along the lines of the above treatment
lead to~\cite{lungo}
\be
Q(t) \simeq {32 \over \pi^2 L^2} \sum_{k=1}^L \exp
\left\{-{\Pi^2 \over 4} \left[k^2+(k+1)^2\right] t \right\}.
\ee 
The sum can be rewritten as the integral 
$\int_1^L \exp[\cdots ] dk$
plus the boundary term for $k=1$: $(1/2)\exp[-(5/4) \Pi^2 t]$
(the boundary term for $k=L$ is always negligible).
It is easy to see that the integral prevails for $t\ll L^2/\pi$
and gives $Q(t)\approx 16 \sqrt{2}/(L \pi^{5/2} \sqrt{t})$,
while in the opposite limit we have an exponential decay.
The existence of the two regimes is clearly shown in Fig.~\ref{fig2}.
The two behaviors can be interpreted physically.
For short times, terrace edges can be neglected and one can focus on the
relative coordinate ($n-m$) of the two particles. Nucleation occurs when
($n-m$) vanishes for the first time and $Q(t) \sim t^{-1/2}$
is simply the spatial integral of the first-passage distribution
probability~\cite{Hughes}. 
The exponential decay for long time is the effect of the
decaying probability that the adatoms remain on the terrace times
the vanishing probability that they have not yet met.

In two dimensions the two diffusing adatoms can be mapped
into a four dimensional problem for a single random walker:
$p_{m_1,n_1,m_2,n_2}(t)$ is the probability
of finding one atom on site $(m_1,n_1)$ and the other in $(m_2,n_2)$
at time $t$. The solution in the noninteracting case on the four
dimensional hypercube is easily found. 
However, the simple generalization of the interacting case to $d=2$ is
not possible because the plane corresponding to
nucleation events does not divide the hypercube into two dynamically separated
regions, reflecting the fact that in two dimensions two adatoms can
exchange their positions without meeting. 
The results shown for the interacting case are obtained by numerically
solving the evolution equation for the probability $p$.

\begin{figure}
\centerline{\psfig{figure=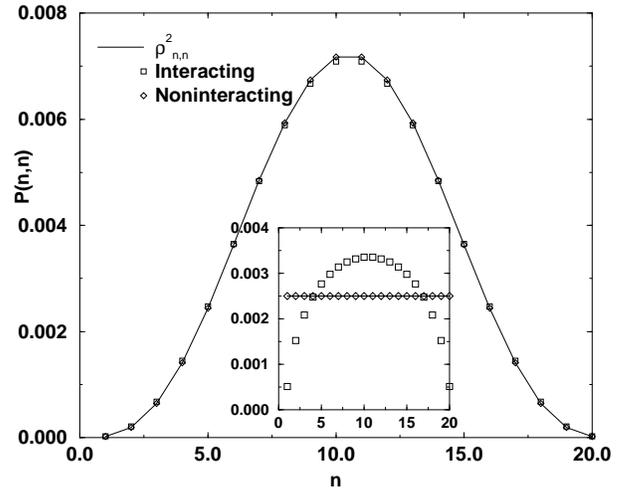,width=8cm,angle=-90}}
\caption{Main: values of the normalized distribution of nucleation sites along
the diagonal $P(n,n)$ for $\les=0$, $d=2$ and $L=20$.
Inset: the same for $\les=\infty$.}
\label{fig3}
\end{figure}

The main part of Fig.~\ref{fig3} clearly shows how the results
obtained in $d=1$ for the  spatial distribution of nucleation sites are also
true in two dimensions:
for non-interacting atoms $P(m,n)\equiv \rho_{m,n}^2$;
in the interacting case a discrepancy exists but is practically negligible.
The distribution of nucleation times $Q(t)$ decays
exponentially for long time, as in $d=1$.
Again, at short time it can be derived~\cite{lungo} by using the
first-passage probability arguments, yielding
--for interacting adatoms-- $Q(t) \sim 1/\ln(t/t_0)$
(Fig.~\ref{fig2}, inset).

Let us now consider $\les=\infty$. In this case, the boundary conditions
for a single adatom in $d=1$ are $p_0(t)=p_1(t)$ and $p_{L+1}(t)=p_L(t)$.
The general solution is now a superposition of the functions
$X_k(n)=\tan[k \pi/(2 L)] \sin(n k \pi/L)+\cos(n k \pi/L)$, with
$k=0,\ldots,L-1$ and the solution of the nucleation problem is found much
in the same way as for zero barriers~\cite{lungo}.
The initial condition is now with both adatoms uniformly distributed,
because for $\les=\infty$ the stationary solution for the single adatom
is $p_n^S=1/L$.

The spatial distribution $P(n)$ of nucleation events is compared to the
mean-field prediction ($P(n)=1/L^d$) in the insets of Fig.~\ref{fig1}
($d=1$) and Fig.~\ref{fig3} ($d=2$): they are in striking disagreement.
We conclude that --concerning $P(n)$-- mean field theory is a good 
approximation for weak step-edge barriers, but the agreement 
is extremely poor for strong ones.

The temporal distribution $Q(t)$ of nucleation events does not change
qualitatively when the value of $\les$ is varied (see Fig.~2).
The power-law ($d=1$) and logarithmic ($d=2$) decays at short time
do not depend on the finite size of the terrace and therefore on $\les$,
which enters in boundary conditions only.
The typical time necessary for the adatom to feel the presence of the
boundaries corresponds to the time $\tau_{tr} \sim L^2/D$.
For longer times $Q(t)$ vanishes exponentially.
In this case the probability for adatoms to remain on the terrace
is clearly constant.
Nonetheless $Q(t)$ {\it must} decay rapidly since its integral
is the total probability of a nucleation event, which is 1 for infinite
barriers: The exponential decay is due to the vanishing
probability that the two adatoms have not yet met before.

>From the experimental point of view, a relevant quantity is the
nucleation rate $\omega$, defined as the number of nucleation events
per unit time on the whole terrace.
Within our framework, we can rigorously show, in all dimensions and for
all values of the barriers, that the mean field value
$\omf= D L^d {\bar \rho}^2$, with ${\bar \rho}= F \tau_{res}$
the average density of occupied sites, must be corrected by a factor
proportional to
${\cal N} \equiv
{N_{\hbox{\scriptsize all}} \over N_{\hbox{\scriptsize dis}}}$, where
$N_{\hbox{\scriptsize dis,all}}$ are, respectively, the number of
{\it distinct} or {\it all} sites visited by a single atom on a terrace.
For reasons of space we only sketch here the derivation of this result,
that will be presented in detail in a longer publication~\cite{lungo}.
The nucleation rate $\omega$ is equal to the deposition rate of a single
particle $1/\tau_{dep}$ times the probability $p_{nuc}$ that such a
particle nucleates a dimer before leaving the terrace.
$p_{nuc}$ is the average over all interarrival times $t'$ of the
probability that two particles meet, when the second is deposited
a time $t'$ after the first.
As discussed previously for the generic quantity $O$, $p_{nuc}$ can be
computed by considering the two particles deposited simultaneously, but
with the first distributed spatially as ${\hat p}_m$. This last function
is shown to be equal to $(\tau_{res}/\tau_{dep}) \; p_m^S$,
where $p_m^S$ is the {\em normalized} solution of the stationary diffusion
equation for a single particle. Hence we have $p_{nuc} = (\tau_{res}/
\tau_{dep}) \;W$, where
$W=\sum_{t=0}^\infty Q(t)$ is the probability that two
adatoms, deposited at the same time with normalized distributions,
meet before leaving the terrace;
$W$ can be shown to be proportional to $N_{\hbox{\scriptsize dis}}/L^d$.
Putting all together and using $\tau_{dep} = 1/(F L^d)$ and
$F \tau_{res}={\bar \rho}$, we obtain
$\omega \sim F L^d {\bar \rho} N_{\hbox{\scriptsize dis}}$.
In a perfectly analogous way one can show, for noninteracting adatoms,
$\oni \sim F L^d {\bar \rho} N_{\hbox{\scriptsize all}}$.
By considering that for noninteracting particles
$\tau_{res} = N_{\hbox{\scriptsize all}}/D$ one sees that $\oni$
is proportional to $\omf$, so that finally
\be
{\omf \over \omega} \sim
{W_{NI} \over W} \sim
{N_{\hbox{\scriptsize all}} \over N_{\hbox{\scriptsize dis}}}
\equiv {\cal N}
\ee
This result holds in all dimensions and for any barrier.

The value of the correction factor ${\cal N}$
depends of course on $d$, $L$ and $\les$.
The numerator $N_{\hbox{\scriptsize all}}$ is just  proportional to
the average density of adatoms on
the terrace: $N_{\hbox{\scriptsize all}}\sim 
L(L+\alpha_d\ell_{\hbox{\tiny ES}})$.
The value of the denominator $N_{\hbox{\scriptsize dis}}$ 
is well known~\cite{Hughes} in
absence of step-edge barriers, being of order $L$ in $d=1$ and of order
$L^2/\ln L$ in $d=2$, and it is trivial in the limit of infinite barriers,
being exactly equal to $L^d$. 
Hence in $d=1$ we obtain ${\cal N}\sim (L+\alpha_1\ell_{\hbox{\tiny ES}})$.
In $d=2$ it is possible to find~\cite{nota2} an
interpolation between the limits ${\cal N}\sim \ln L$ ($\les\ll L$)
and ${\cal N}\sim \les/L$ ($\les\gg L$):
${\cal N}\sim (1+\alpha_2\ell_{\hbox{\tiny ES}}/L)/p_s$,
where $p_s = 1-[1-1/(\ln L)]^{\tau_{res}/\tau_{tr}}$.
The previous interpolation is in reasonable agreement with 
Monte Carlo simulations~\cite{lungo} performed for intermediate barriers.

In conclusion we have provided exact results for the spatio-temporal
distributions and total nucleation rates for irreversible nucleation
on terraces during crystal growth. These results should be used 
wherever mean-field approximate results were so far commonly used.
One example is the experimental determination of ES barriers from the
rate of nucleation on terraces~\cite{TDT,altro}.
Another example is the modelization of epitaxial growth with mesoscopic
models~\cite{EV}:
the location of new terraces must be chosen according to the
correct spatial distribution of nucleation events, derived here.

We thank S. Redner for pointing out useful references and
F. Colomo and E. Caglioti for interesting discussions.


\begin{references}
\bibitem{nota1}
It is sufficient to sum on $t$ the two sides of 
Eq.~(\protect\ref{evolution_1d}).
\bibitem{nota2}
The argument is as follows.
The atom performs on average a number $N_{tr}=\tau_{res}/\tau_{tr}$
of traversals of the island.  During a single traversal each site
has a probability $p_1\sim (1/\ln L)$ to be visited. After all 
$N_{tr}$ traversals the probability $p_s$ that a generic site has been
visited at least once is given by $1-p_s = (1-p_1)^{N_{tr}}$.
\end{references}
\end{document}